\documentstyle[erice98,12pt,psfig]{article}

\newcommand{\be}[1]{\begin{equation} \label{(#1)}}
\newcommand{\ee}{\end{equation}}
\newcommand{\ba}[1]{\begin{eqnarray} \label{(#1)}}
\newcommand{\ea}{\end{eqnarray}}

\begin{document}
\title{Effective kaon energy from a novel chiral SU(3) model}
\author{Guangjun Mao$^{1}$, P. Papazoglou$^{1}$, S. Hofmann$^{1}$,
S. Schramm$^{2}$,
H.~St\"{o}cker$^{1}$, and W.~Greiner$^{1}$}
\address{$^{1}$Institut f\"{u}r Theoretische Physik der
J. W. Goethe-Universit\"{a}t \\
Postfach 11 19 32,  D-60054 Frankfurt am Main, Germany \\
$^{2}$GSI Darmstadt, Postfach 11 05 52, D-64220 Darmstadt, Germany}
\maketitle
\vspace{-0.25cm}
\abstracts{A new chiral SU(3) Lagrangian is proposed         
to describe the properties of  
kaons and anti-kaons in the nuclear medium.        
The saturation properties of nuclear matter are reproduced
as well as the results
of the Dirac-Br\"{u}ckner theory. After introducing the coupling between the
omega meson and the kaon, our results for effective
 kaon and anti-kaon energy are quite similar as 
calculated in the one-boson-exchange model. }
\vspace{-0.3cm}
\begin{sloppypar}
The properties of kaons/anti-kaons in nuclear and neutron matter have
attracted considerable interest. 
Substantial theoretical effort 
has  been
devoted to investigate medium effects on  kaons 
and anti-kaons in dense 
matter \cite{Bro92,Sch94,Koc94,Waa96}. Recent data by 
the Kaos collaboration \cite{Bar97} on $K^{+}$ and $K^{-}$ production in 
relativistic heavy-ion collisions,
which seems to exhibit
a substantial enhancement of the $K^{-}$ yield, stimulated 
further activity. Among the various  models proposed, the chiral   
SU(3) Lagrangian seems to be particularly useful, since the 
kaon is  essentially a pseudo-Goldstone
boson.  However, up to now, a  consistent calculation based on a chiral 
Lagrangian, which can simultaneously describe both 
the kaon-nuclear interactions and the ground state of the dense matter,
has not been performed yet. 

\end{sloppypar}
\begin{sloppypar}
This is the aim of the present work.
It addresses the problem in a novel chiral SU(3)-symmetric Lagrangian
\cite{Pap98}. 
In addition to the ground-state saturation properties of 
nuclear matter, the whole density dependence of the mean fields as  
predicted by the Dirac-Br\"{u}ckner theory \cite{Bro90} are considered 
as a further 
constraint to the model. This will turn out to be rather
important for the investigation of the kaon
and anti-kaon properties at higher densities.
The chiral SU(3) Lagrangian reads
\begin{equation}
{\cal L} = {\cal L}_{kin} + \sum_{W=X,V,u,\Gamma}{\cal L}_{BW}
+ {\cal L}_{vec} + {\cal L}_{0} + {\cal L}_{SB}.
\end{equation}
The main feature of the model is that the baryon masses are generated by the
scalar condensates while their splitting is realized through SU(3) symmetry
breaking for these condensates.
The model is described in  \cite{Pap98}. 
Considering SU(3) generators up to quadratic order, one can obtain the 
Lagrangian for nuclear matter ground state and kaon-nuclear interaction
simultaneously \cite{Pap98,Mao}.  
The kaon interaction is described by
\begin{equation}
{\cal L}_{KN} = - \frac{3i}{8f_{K}^{2}}\bar{\psi}\gamma_{\mu}\psi
\bar{K}\stackrel{\leftrightarrow}{\partial^{\mu}}K + \frac{m_{K}^{2}}
{2f_{K}} \left( \sigma + \sqrt{2}\zeta \right) \bar{K}K
-i{\rm g}_{\omega K}\bar{K}\stackrel{\leftrightarrow}{\partial^{\mu}}K
\omega_{\mu}.
\end{equation}
The omega-kaon coupling is introduced through considering the vector field as
a gauge field.  
The parameters of the model are determined by the SU(3) vacuum and
the saturation properties of nuclear matter. 
The corresponding saturation properties are:
$m^{*}/m_{N}=0.641$, $E/A(\rho_{0})=-15.93$ MeV, $K=266.2$ MeV
and $\rho_{0}=0.15$ $fm^{-3}$;   
$f_{\pi}=93.3$ MeV, $f_{K}=122$ MeV, $m_{\pi}=139$ MeV, $m_{K}=498$ MeV.
The parameters of the kaon-nuclear interactions,  
Eq. (2),
are constrained by the chiral Lagrangian itself.  
The model can reproduce the results of the
G-matrix calculations \cite{Bro90} up to four times normal density \cite{Mao}. 
After making a field shift to new variables $\phi$ and $\xi$
($\sigma=\sigma_{0} - \phi$, $\zeta = 
\zeta_{0} - \xi$), we obtain 
the effective-mass and -energy of the kaon $K$ 
and the anti-kaon $\bar{K}$
\begin{eqnarray}
&& m^{*2}_{K} = m_{K}^{2} + \frac{m_{K}^{2}}{2f_{K}}\phi 
+\frac{m_{K}^{2}}{\sqrt{2}f_{K}}\xi , \\
&& \omega_{K,\bar{K}}=  \sqrt{ m^{*2}_{K}
+ \left( \frac{3}{8f_{K}^{2}}\rho_{B} + {\rm g}_{\omega K}\omega_{0}\right) ^{2} }
\pm \left( \frac{3}{8f_{K}^{2}}\rho_{B} + {\rm g}_{\omega K}\omega_{0}
\right).
\end{eqnarray}
\noindent Fig.~1 depicts the kaon and anti-kaon effective energies calculated 
with Eqs. (3) and (4). The results of other models are also displayed for
comparison. It can be seen that without the $\omega - K$ coupling our model
predicts a rather weak potential for the anti-kaon compared to the
predictions of other models. After introducing the
$\omega - K$ coupling, the calculated effective energies for kaon and
anti-kaon are quite similar as obtained in the one-boson-exchange model 
\cite{Sch94}. 
At normal density the optical potentials of kaon and anti-kaon are
$U_{opt}^{K}=21.53$ MeV and $U_{opt}^{\bar{K}}=-98.7$ MeV, respectively,
which are in accordance with the data from the $KN$ scattering lengths
 and the results of other models \cite{Sch94,Waa96}. However,
the $U_{opt}^{\bar{K}}$ is still much weaker than the prediction from the
$K^{-}$ atomic data \cite{Fri93}, which gives $U_{opt}^{\bar{K}}=-200\pm20$
MeV. Further investigation is needed.

\end{sloppypar}
\vspace{-0.8cm}
\begin{minipage}{8.0cm}
\hspace{-0.5cm}
 \centerline{\psfig{figure=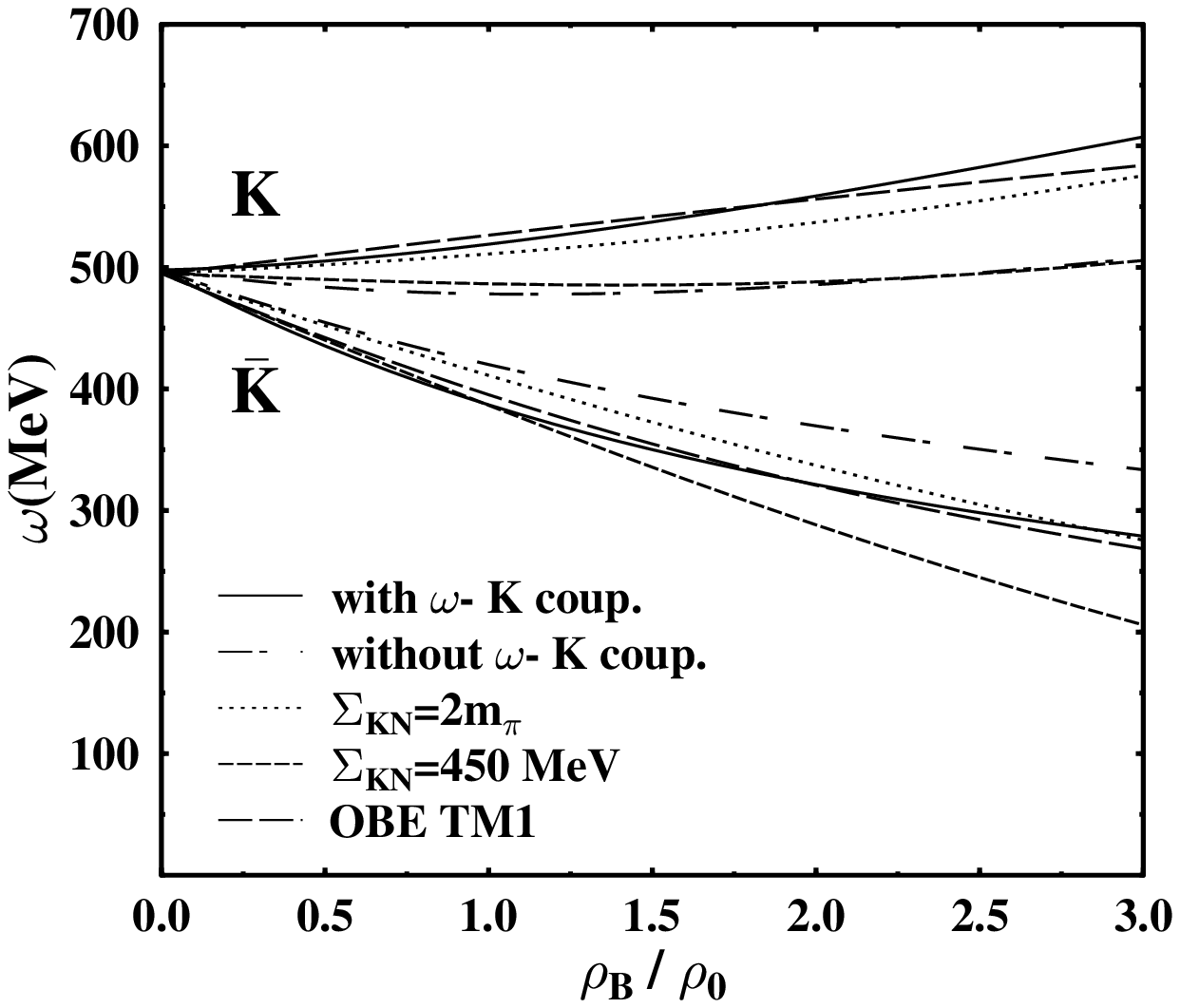,width=8.0cm,height=9.0cm,angle=-90}}
 \end{minipage}\hfill
\begin{minipage}{8.0cm}
{\footnotesize Fig.~1: The energies of kaons and anti-kaons 
as a function of the density. 
The solid
and dot-dashed line represent the results of this work with and without the
$\omega - K$ coupling, respectively. The dotted and short dashed line are
calculated with chiral perturbation theory with the different $KN$ sigma terms.
The long dashed line depicts the results of one-boson-exchange model with the
TM1 parameter set.}
 \end{minipage}
\vspace{-0.5cm}
\renewcommand{\baselinestretch}{1.0}

{\footnotesize
\begin{thebibliography}{250}
\bibitem{Bro92}
   G.E.~Brown et al.,                      
   Z. Phys. A341 (1992) 301;
   Nucl. Phys. A567 (1994) 937.
\bibitem{Sch94}
   J.~Schaffner et al.,                                                   
   Phys. Lett. B334 (1994) 268;
   Phys. Rev. C53 (1996) 1416;
   Nucl. Phys. A625 (1997) 325.
\bibitem{Koc94}
   V.~Koch
   Phys. Lett. B337 (1994) 7.
\bibitem{Waa96}
   T.~Waas et al.,               
   Phys. Lett. B365 (1996) 12;
   Phys. Lett. B379 (1996) 34;
   Nucl. Phys. A594 (1995) 325.           
\bibitem{Bar97}
   R.~Barth et al.,                     
   Phys. Rev. Lett. {\bf 78} (1997) 4007.
\bibitem{Pap98}
   P.~Papazoglou et al., 
   Phys. Rev. C57 (1998) 2576;
   nucl-th/9806087, Phys. Rev. C in press.
\bibitem{Bro90}
   R.~Brockmann and R.~Machleidt,
   Phys. Rev. C42 (1990) 1965.
\bibitem{Mao}
   Guangjun~Mao et al., nucl-th/9806068.
\bibitem{Fri93}
   E.~Friedman, A.~Gal, and C.J.~Batty,
   Phys. Lett. B308 (1993) 6;
   Nucl. Phys. A579 (1994) 518.
\end{thebibliography}}
\end{document}